\documentclass[11pt]{article}
\usepackage{epsf}
\usepackage{graphicx}
\newlength{\bredde}
\def\slash#1{\settowidth{\bredde}{$#1$}\ifmmode\,\raisebox{.15ex}{/}
\hspace*{-\bredde} #1\else$\,\raisebox{.15ex}{/}\hspace*{-\bredde} #1$\fi}
\newcommand{\beq}{\begin{equation}}
\newcommand{\eeq}{\end{equation}}
\newcommand{\noi}{\vspace{12pt}\noindent}
\newcommand{\lG}{\raise.3ex\hbox{$\stackrel{\leftarrow}{G}$}}
\newcommand{\lU}{\raise.3ex\hbox{$\stackrel{\leftarrow}{U}$}}
\newcommand{\lP}{\raise.3ex\hbox{$\stackrel{\leftarrow}{{\cal P}}$}}
\newcommand{\leta}{\raise.3ex\hbox{$\stackrel{\leftarrow}{\eta}$}}
\newcommand{\lOmega}{\raise.3ex\hbox{$\stackrel{\leftarrow}{\Omega}$}}
\newcommand{\ldr}{\raise.3ex\hbox{$\stackrel{\leftarrow}{\delta^r}$}}

\def\beqn{\begin{eqnarray}}
\def\eeqn{\end{eqnarray}}

\def\gtwid{\raise.3ex\hbox{$>$\kern-.75em\lower1ex\hbox{$\sim$}}}
\def\ltwid{\raise.3ex\hbox{$<$\kern-.75em\lower1ex\hbox{$\sim$}}}

\begin{document}
\topmargin -1.4cm
\oddsidemargin -0.8cm
\evensidemargin -0.8cm
\title{\Large{{\bf Eigenvalue Distributions of the QCD Dirac Operator }}}

\vspace{1.5cm}

\author{~\\~\\
{\sc P.H. Damgaard$^a$, U.M. Heller$^b$, R. Niclasen$^a$} and 
{\sc K. Rummukainen$^{c,d}$}\\~\\~\\
$^a$The Niels Bohr Institute and $^c$NORDITA\\ Blegdamsvej 17\\ 
DK-2100 Copenhagen, Denmark\\~\\~\\
$^b$CSIT\\Florida State University\\Tallahassee, FL 32306-4120, USA\\~\\~\\
$^d$Helsinki Institute of Physics,\\
P.O.Box 9, 00014 University of Helsinki, Finland\\}
\maketitle
\vfill
\begin{abstract} 
We compute by Monte Carlo methods the individual distributions of the $k$th
smallest Dirac operator eigenvalues in QCD, and compare them with recent
analytical predictions. We do this for both massless and massive
quarks in an SU(3) gauge theory with staggered fermions. Very
precise agreement is found in all cases. As a simple by-product we
also extract the microscopic spectral density
of the Dirac operator in SU(3) gauge theory with dynamical massive fermions
for $N_f=1$ and 2, 
and obtain high-accuracy agreement with analytical expressions.
\end{abstract}
\vfill
\begin{flushleft}
NBI-HE-00-34 \\
CSIT2000-17\\
NORDITA-2000/61 HE\\
hep-lat/0007041
\end{flushleft}
\thispagestyle{empty}
\newpage

\setcounter{page}{1}

\noindent
In the chiral limit of QCD it is possible to derive exact analytical
expressions for eigenvalue distributions and eigenvalue correlations
of the Dirac operator. This situation arises if one restricts oneself
to the phase of spontaneously broken chiral
symmetry (with infinite-volume zero-mass condensate $\Sigma$), and considers
the theory in a finite space-time volume $V$. Taking the chiral
limit by sending quark masses $m_i$ to zero in such a way that the
combination $\mu_i \equiv m_i\Sigma V$ is kept fixed as $V \to \infty$,
this leads to an interesting finite-volume scaling regime of QCD. 
Denoting generically all pseudo-Goldstone masses by 
$m_{\pi}$, exact scaling is achieved if $m_{\pi} \ll V^{-1/4}$ \cite{LS}.  
The only assumption is that chiral symmetry is spontaneously broken
from one group $G$ to a smaller group $H$. The coset $G/H$
specifies the universality class \cite{SV}.

\noi
The most detailed analytical predictions were originally obtained from
an intriguing exact relation to universal Random Matrix Theory results
\cite{SV,ADMN}. Later, the same results have been derived
directly from the effective partition functions \cite{AD} based 
on the so-called supersymmetric formulation \cite{OTV}. Very
recently, also the replica method \cite{DS} has been shown to
yield identical results \cite{DV}. Beyond any doubt now, 
these analytical predictions represent exact statements about the Dirac 
operator spectrum of QCD in the above finite-volume scaling region. 
This is an exciting situation from the point of view of lattice gauge 
theory, which is almost tailored to study such a finite-volume regime.
It is one of the few instances where there are exact and non-perturbative
analytical predictions that immediately can be compared with lattice
Monte Carlo results. There have by now been several Monte Carlo
studies of these analytical predictions for Dirac operator eigenvalues
in various four-dimensional gauge theories
\cite{BBetal,su2mass,DHK,EHN_stag,EHN,Berg}.
One peculiarity of these predictions is that they are most easily 
expressed in gauge field sectors of fixed topological charge 
$\nu$\footnote{The sum over topology can be performed \cite{top}, but closed
analytical forms are not known except in very simple cases. For instance,
in theories with massless (dynamical) fermions only the sector with 
$\nu=0$ contributes.}.
For fermions sensitive to topology this gives an interesting new possibility,
as predictions differ markedly for different topological sectors \cite{EHN}.
But staggered fermions seem oblivious to gauge field topology at the
gauge couplings that are presently realistic \cite{DHNR,Lang}; predictions
can then only be made with respect to the topologically trivial sector of
$\nu=0$.

\noi
In this letter we shall extend the Monte Carlo analyses of Dirac operator
eigenvalues in two directions. First, we shall confront some very recent
analytical results concerning the probability distribution of the
$k$th smallest Dirac operator eigenvalue \cite{ksmall} with Monte Carlo
data. This is important, because there is much more detailed information
in these individual eigenvalue distributions than in, for example, the
spectral density, which is the {\em sum} of all these individual
distributions (see below).
Second, we shall make the extension of the quenched results
of Ref.~\cite{DHK} to the case of dynamical fermions of non-zero mass.
Until now, Monte Carlo analyses of microscopic Dirac operator spectra 
with dynamical fermions have been limited to a different
universality class\footnote{An SU(2) gauge theory labeled by the 
so-called ``Dyson index'' $\beta=4$ for staggered fermions \cite{SV}.} 
than the one relevant for QCD \cite{su2mass} (see also Ref.~\cite{AK}). 
We study gauge group SU(3), which, even with staggered
fermions, allows us to compare with the universality class relevant
for continuum QCD (the class of Dyson index $\beta=2$). 

\noi
Details of how we compute the lowest-lying Dirac operator eigenvalues
have been given in Ref.~\cite{DHNR}, and we shall not repeat them
here. Instead, let us briefly recall the analytical predictions with
which we should compare our Monte Carlo data. We start with the recently
derived expression for the distribution of the $k$th Dirac operator
eigenvalue in the finite-volume scaling region mentioned above. The
generator for these individual eigenvalue distributions is the joint 
probability distribution for the $k$ smallest eigenvalues. It turns out
that this joint probability distribution can be written very compactly
in terms of finite-volume partition functions with additional fermion
species. For the universality class relevant for QCD, the explicit
expression is \cite{ksmall}:
\begin{eqnarray}
&&\omega_k(\zeta_1,\ldots,\zeta_k; \{\mu\})
~=~ 
C~ e^{-\zeta_k^2/4}
\zeta_k
\prod_{i=1}^{k-1}\Bigl(\zeta_i^{2\nu+1}
\prod_{j=1}^{N_{f}}(\zeta_i^2+\mu_j^2)
\Bigr)\prod_{i>j}^{k-1}(\zeta_i^2-\zeta_j^2)^2
\prod_{j=1}^{N_{f}}\mu_j^{\nu} \cr
&&\times
\frac{{\cal Z}_{2}\Bigl(
\left\{\sqrt{\mu_i^2+\zeta_k^2}\right\},
\sqrt{\zeta_k^2-\zeta_1^2},\sqrt{\zeta_k^2-\zeta_1^2},
\ldots,
\sqrt{\zeta_k^2-\zeta_{k-1}^2},\sqrt{\zeta_k^2-\zeta_{k-1}^2},
\overbrace{\zeta_k,\ldots,\zeta_k}^{\nu}\Bigr)}{
{\cal Z}_{\nu}(\{\mu\})} ~.
\label{bboxOmega}
\end{eqnarray}
Here the additional $2(k-1)$ 
fermion species of masses $\sqrt{\zeta_k^2-\zeta_i^2}$
are doubly degenerate, as indicated. The partition function in the
numerator is evaluated in a sector of {\em fixed} topological charge
$\nu=2$ (independent of the actual topological charge in question). But
there is information about $\nu$ in the fact that $\nu$ additional
fermion species (all degenerate in mass $\zeta_k$) 
enter in the partition function.
Actually, since we shall be dealing with staggered fermions in
the present paper, only the sector with $\nu=0$ is relevant. In that
case the above formula simplifies:
\begin{eqnarray}
&&\omega_k(\zeta_1,\ldots,\zeta_k; \{\mu\})
~=~ 
C~ e^{-\zeta_k^2/4}
\zeta_k
\prod_{i=1}^{k-1}\Bigl(\zeta_i
\prod_{j=1}^{N_{f}}(\zeta_i^2+\mu_j^2)
\Bigr)\prod_{i>j}^{k-1}(\zeta_i^2-\zeta_j^2)^2\cr
&&\times
\frac{{\cal Z}_{2}\left(
\left\{\sqrt{\mu_i^2+\zeta_k^2}\right\},
\sqrt{\zeta_k^2-\zeta_1^2},\sqrt{\zeta_k^2-\zeta_1^2},
\ldots,
\sqrt{\zeta_k^2-\zeta_{k-1}^2},\sqrt{\zeta_k^2-\zeta_{k-1}^2}\right)}{
{\cal Z}_{0}(\{\mu\})} ~.
\label{bboxOmega1}
\end{eqnarray}

\noi
The fact that in particular the smallest (non-zero) Dirac operator eigenvalue
has a distribution determined entirely in terms of the effective
partition function was noted earlier \cite{NDW}, but the present
most general expression is actually more compact. The proportionality
factor $C$ depends on the chosen normalization of the partition
functions. We choose a normalization in which \cite{JSV,AD}
\beq
{\cal Z}_{\nu}(\{\mu\}) ~=~ \det A(\{\mu\})/\Delta(\mu^2) ~,
\eeq
where the determinant is taken over the $N_f\times N_f$ matrix
\beq
A(\{\mu\}) ~=~ \mu_i^{j-1}I_{\nu+j-1}(\mu_i) ~,
\eeq
and 
\beq
\Delta(\mu^2) ~=~ \prod_{i>j}^{N_{f}}(\mu_i^2-\mu_j^2) ~.
\eeq
With this convention the normalization factor is $C = 1/2$ for all
values of $k$, $N_f$ and $\nu$.

\noi
For the purpose of comparing with lattice gauge theory data, a more
convenient quantity to focus on is the probability distribution of the
$k$th smallest Dirac operator eigenvalue \cite{ksmall}. One gets it
from the joint probability distribution by integrating out the 
previous $k-1$ smaller eigenvalues: 
\beqn
{p}_{k}(\zeta;\{\mu\}) &=&
\int_0^\zeta d\zeta_1 \int_{\zeta_{1}}^\zeta d\zeta_2 
\cdots \int_{\zeta_{k-2}}^\zeta d\zeta_{k-1}
~\omega_{k}(\zeta_1,\ldots,\zeta_{k-1},\zeta;\{\mu\}) \cr
&=& \frac{1}{(k-1)!}
\int_0^\zeta d\zeta_1 \int_{0}^\zeta d\zeta_2 
\cdots \int_{0}^\zeta d\zeta_{k-1}
~\omega_{k}(\zeta_1,\ldots,\zeta_{k-1},\zeta;\{\mu\}) ~.\label{p_k}
\eeqn

\noi
The individual eigenvalue distributions by definition sum up to the
microscopic spectral density $\rho_S(\zeta;\{\mu\})$:
\beq
\rho_S(\zeta;\{\mu\}) ~=~ \sum_{k=1}^\infty
{p}_{k}(\zeta;\{\mu\}).
\eeq
But a much simpler way to get the microscopic spectral density is by
direct evaluation. As we will in addition 
be interested in the case of dynamical
fermions with non-zero quark masses $\mu_i$, we need the microscopic
spectral density for the general massive case of this universality class
\cite{mass0,mass,mass1}.
Also this quantity can most simply be expressed
in terms of the finite-volume partition functions \cite{AD}:
\beq\label{eq:rhoS}
\rho_S(\zeta;\{\mu\}) ~=~ (-1)^{\nu} |\zeta|\prod_{j=1}^{N_{f}}
(\zeta^2+\mu_j^2)\frac{{\cal Z}_{\nu}(\{\mu\},i\zeta,i\zeta)}{
{\cal Z}_{\nu}(\{\mu\})} ~,
\eeq
using the same normalization convention as above.
This spectral density is ``double-microscopic'' in the sense that
both eigenvalues $\lambda$ and masses $m_j$ have been rescaled according
to $\zeta =\lambda\Sigma V$ and $\mu_j = m_j\Sigma V$. As the rescaled
masses $\mu_j$ are sent to infinity, one obtains a series of decoupling
relations where the corresponding quark flavors become effectively
quenched \cite{mass}. Because we precisely want to see the effects of having
dynamical fermions, it is thus important to keep the rescaled masses
$\mu_i$ on roughly the same scale as the eigenvalues, or smaller.

\begin{figure}
\center
\includegraphics[scale=0.7]{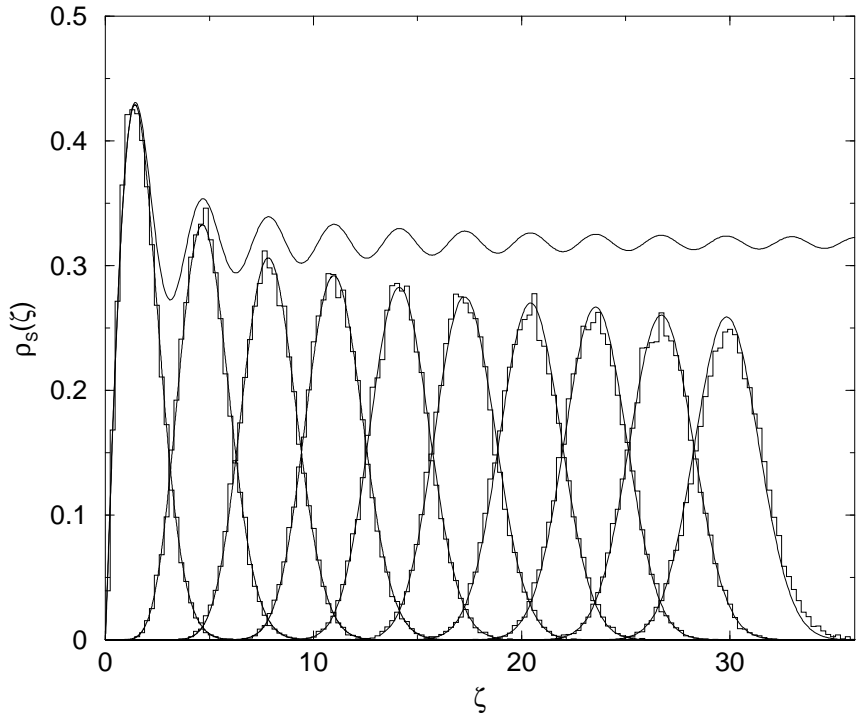}
\caption{Simulation data of the 10 lowest quenched Dirac operator eigenvalues
compared with the analytic predictions for the individual
eigenvalues. The data were obtained on an $8^4$ lattice at $\beta=5.1$.}
\label{fig:quenched}
\end{figure}

\noi 
We begin by comparing {\em quenched} lattice gauge theory data with
the predictions for the individual eigenvalue distributions. We have
here large statistics for the 10 smallest Dirac operator
eigenvalues on an $8^4$ lattice at coupling $\beta = 5.1$. These
parameters and all others used in this paper are summarized in
Table~\ref{tab:param}. In Fig.~\ref{fig:quenched} we show a comparison of
these lattice Monte Carlo results with the formula (\ref{p_k}) for
$k=1\ldots10$. $\Sigma$ was determined from a fit of the distribution
of the lowest eigenvalue.
The formula (\ref{p_k}) obviously gets quite tedious
to evaluate by numerical integration for large values of $k$. 
The analytic curves for the distributions of the last
couple of eigenvalues have actually not been obtained directly from 
(\ref{p_k}), but rather from the eigenvalue distributions of computer 
generated random matrices (which obviously amounts to the same). 
The agreement with our lattice Monte Carlo data is nothing less than 
spectacular for all the
eigenvalues we have available.  Eventually, on any given finite
lattice, $1/V$-corrections will take over, and this very precise
agreement with the theoretical predictions will no longer hold. But in
the limit $V \to \infty$ the agreement should persist for an infinity
of smallest Dirac operator eigenvalues $\lambda_n$, rescaled so that
$\zeta_n \equiv \lambda_n \Sigma V$ remains finite in the
infinite-volume limit. Agreement with these lattice data and the
microscopic spectral density has already been presented in
Ref.~\cite{DHNR}, but after just a few oscillations the spectral
density quickly becomes an almost constant function, with very little
information. For the first time we see here that there is detailed
agreement with the analytical predictions for a whole sequence of
individual eigenvalues underneath.

\begin{table}
\begin{center}
\begin{tabular}{c|c|c|c|c|c}
$V$ & N$_f$ & $\beta$ & $\Sigma$ & $m$ & \#configs \\
\hline 
$4^4$ & 1 & 4.6 & 1.24 & 0.02 & 40,600 \\
& & & & 0.01 & 32,650 \\
& & & & 0.004 & 25,800 \\
\hline 
$4^4$ & 2 & 4.2 & 1.28 & 0.01 & 26,040 \\
\hline 
$6^4$ & 1 & 4.7 & 1.17 & 0.003 & 3,370 \\
\hline 
$8^4$ & 0 & 5.1 & 1.15 & -  & 17,454 \\
\hline
\end{tabular}
\end{center}
\caption{Simulation parameters. Note that the infinite-volume zero-mass chiral
condensate $\Sigma$ is one single free parameter, depending only on
$\beta$ and $N_f$. For the case of $V=4^4$ and $N_f=1$ there are thus
no free parameters at all for two of three simulations with different
quark masses.}
\label{tab:param}
\end{table}

\noi 
We next turn to simulations with dynamical fermions of finite mass.
As mentioned above, there has previously been only one Monte Carlo
study in this direction \cite{su2mass}, for a different gauge
group (and hence different universality class) SU(2). We shall here
provide first results for a study with gauge group SU(3), and hence
the universality class of continuum QCD. We shall not only compare
with the microscopic spectral density, but also focus on individual
eigenvalue distributions, just as we did above in the quenched case.

\noi
For our dynamical simulations we employed the Hybrid Monte Carlo
algorithm with $N_f = 1$ or 2 species of staggered fermions. In the
continuum limit, this would correspond to $n_f = 4 N_f$ continuum
fermion flavors. However, in order to have a large physical volume
with few lattice sites --- our computer resources limit us to the
use of systems with $4^4$ up to $6^4$ lattice sites and the finite
size scaling limit we want to consider requires a sufficiently large
physical volume, $V \gg \Lambda_{QCD}^{-4}$ ---
we are forced to work at a large lattice spacing
and hence a large gauge coupling. In such a situation the flavor
symmetry of the staggered fermions is badly broken, especially in the
limit of small masses of interest here. Only the $N_f^2$ true 
pseudo-Goldstone pions from the spontaneous breaking of the
$U(N_f) \times U(N_f)$ symmetry to the diagonal $U(N_f)$ symmetry
are light. The flavor symmetry breaking makes the other $n_f^2 - 1 - N_f^2$
``would-be'' pseudo-Goldstone pions much heavier. As far as the infrared limit
is concerned, only $N_f$ flavors contribute. This will be verified by
our results presented below.

\begin{figure}
\begin{tabular}{c c c}
\includegraphics[scale=0.33]{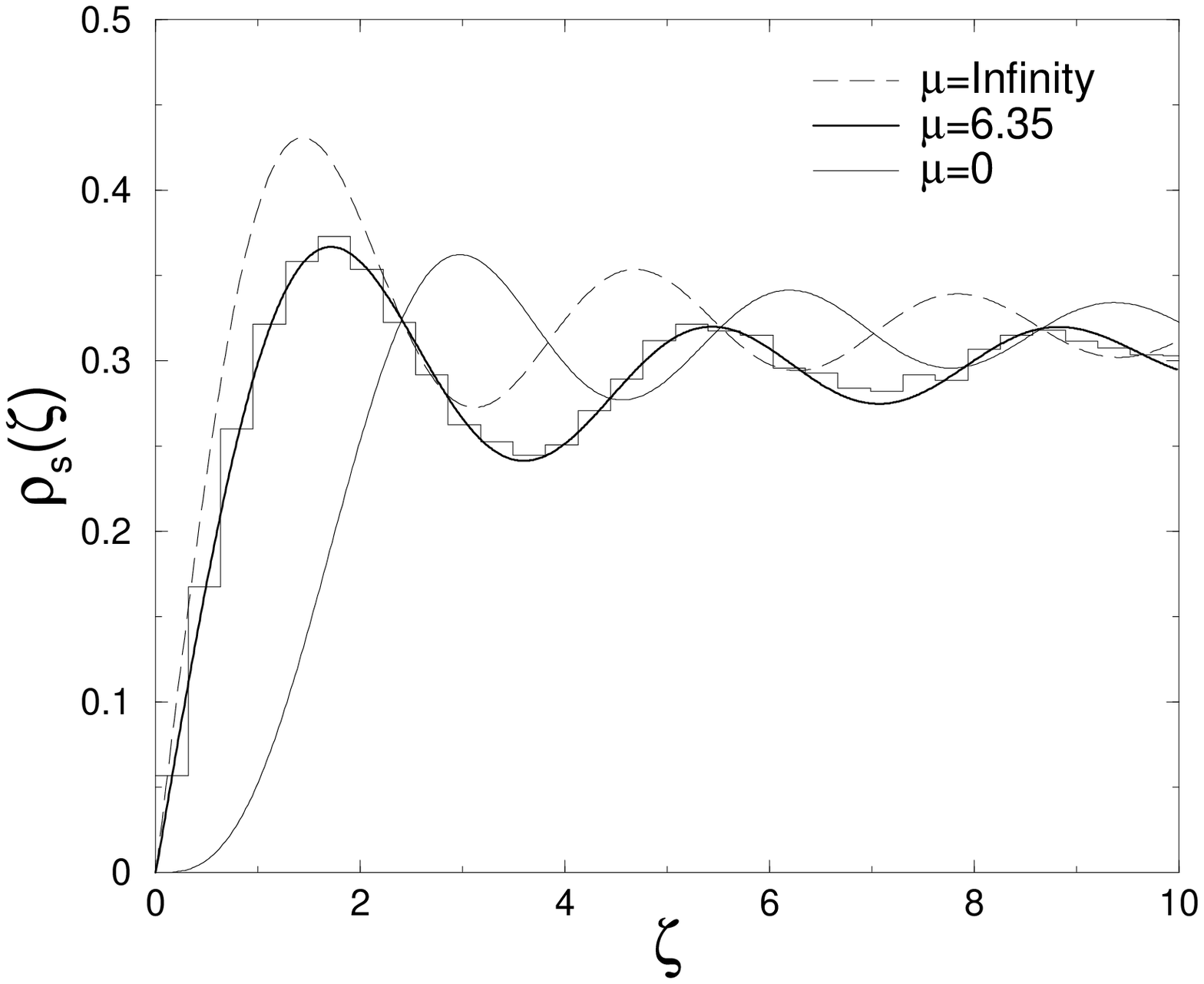}
&
\includegraphics[scale=0.33]{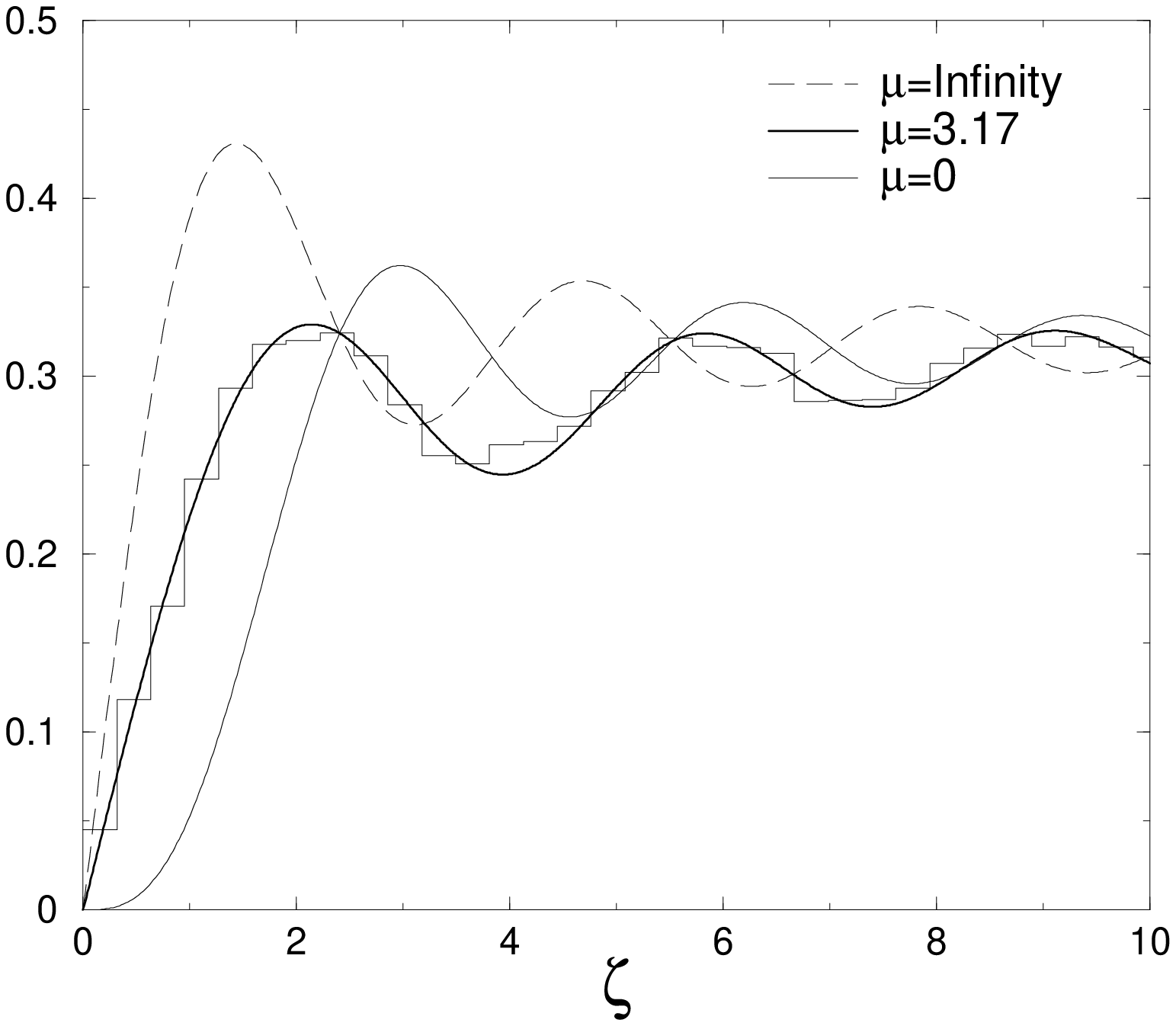}
&
\includegraphics[scale=0.33]{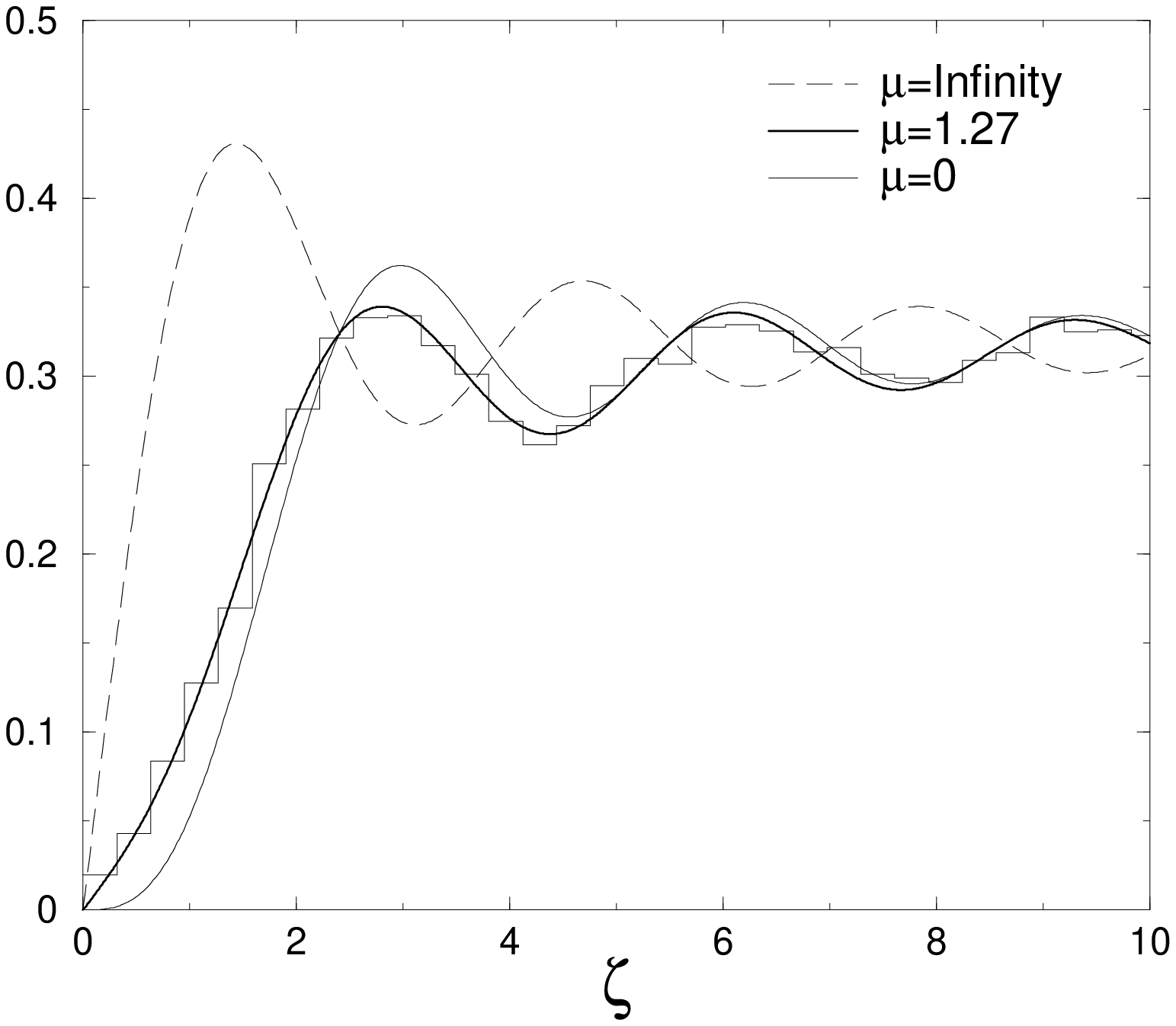}
\end{tabular}
\caption{The microscopic spectral density for different quark masses on a 
$4^4$ lattice with $N_f=1$ at $\beta=4.6$. 
The three simulation masses $m= 0.02, 0.01$
and 0.004 correspond to $\mu = 6.35, 3.17$ and 1.27, respectively.
Also the curve for one massless flavor and 
the N$_f$=0 (quenched) curve, which is equivalent to the limit
$\mu \to \infty$, is plotted. There is excellent agreement with the analytical
curves corresponding to just the right $\mu$-value.}
\label{fig:dyn_rho}
\end{figure}

\begin{figure}
\begin{tabular}{c c c}
\includegraphics[scale=0.33]{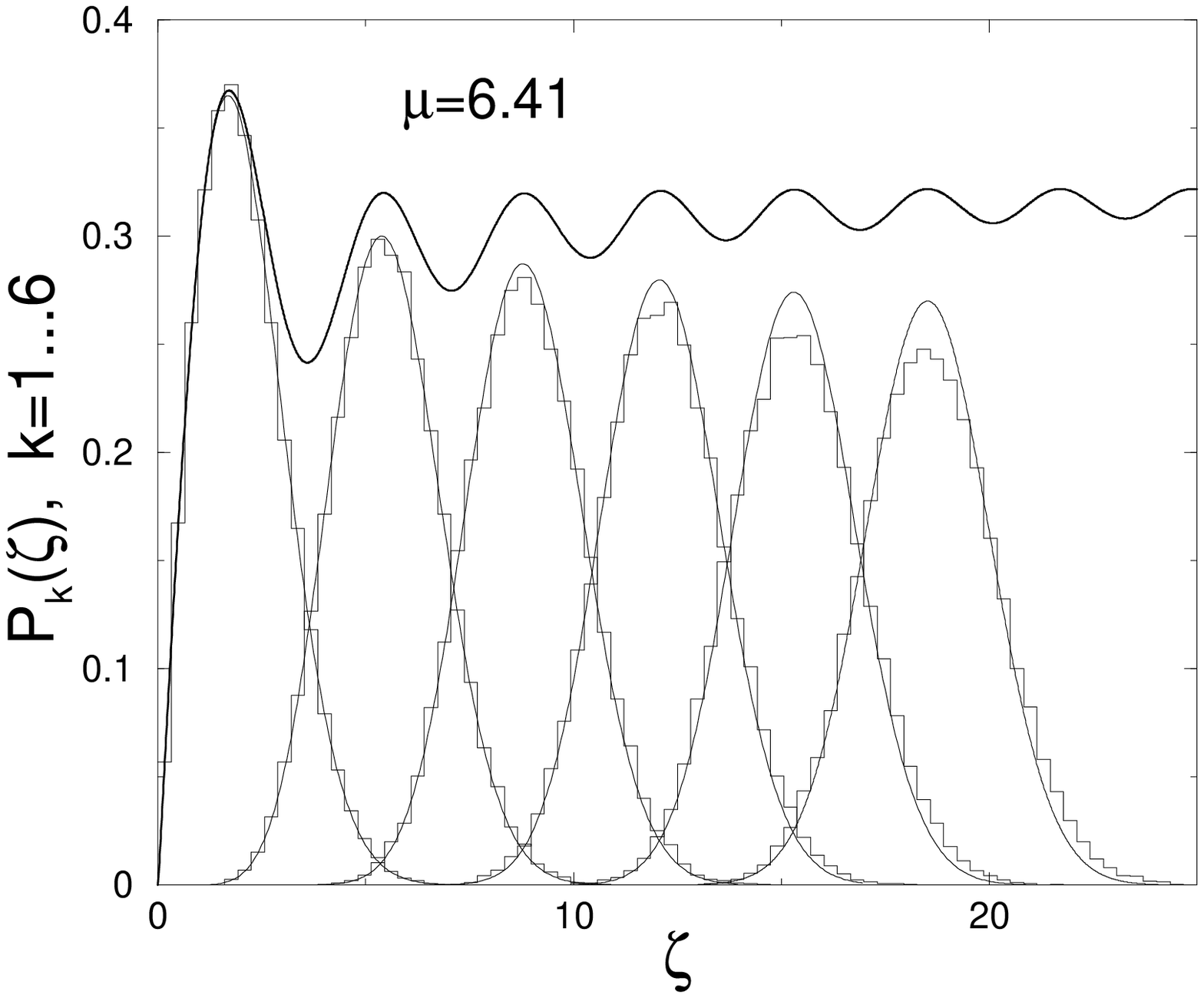}
&
\includegraphics[scale=0.33]{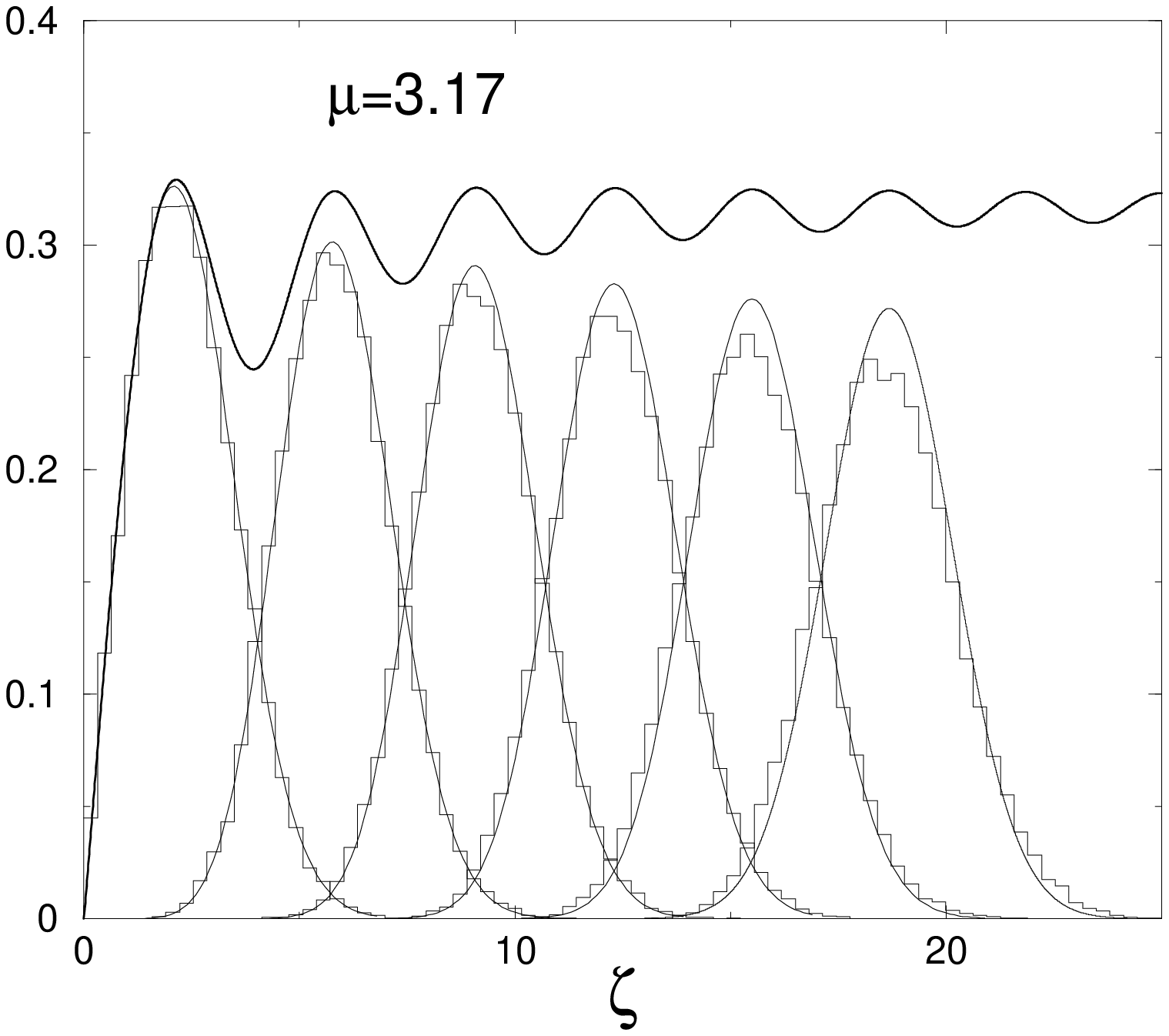}
&
\includegraphics[scale=0.33]{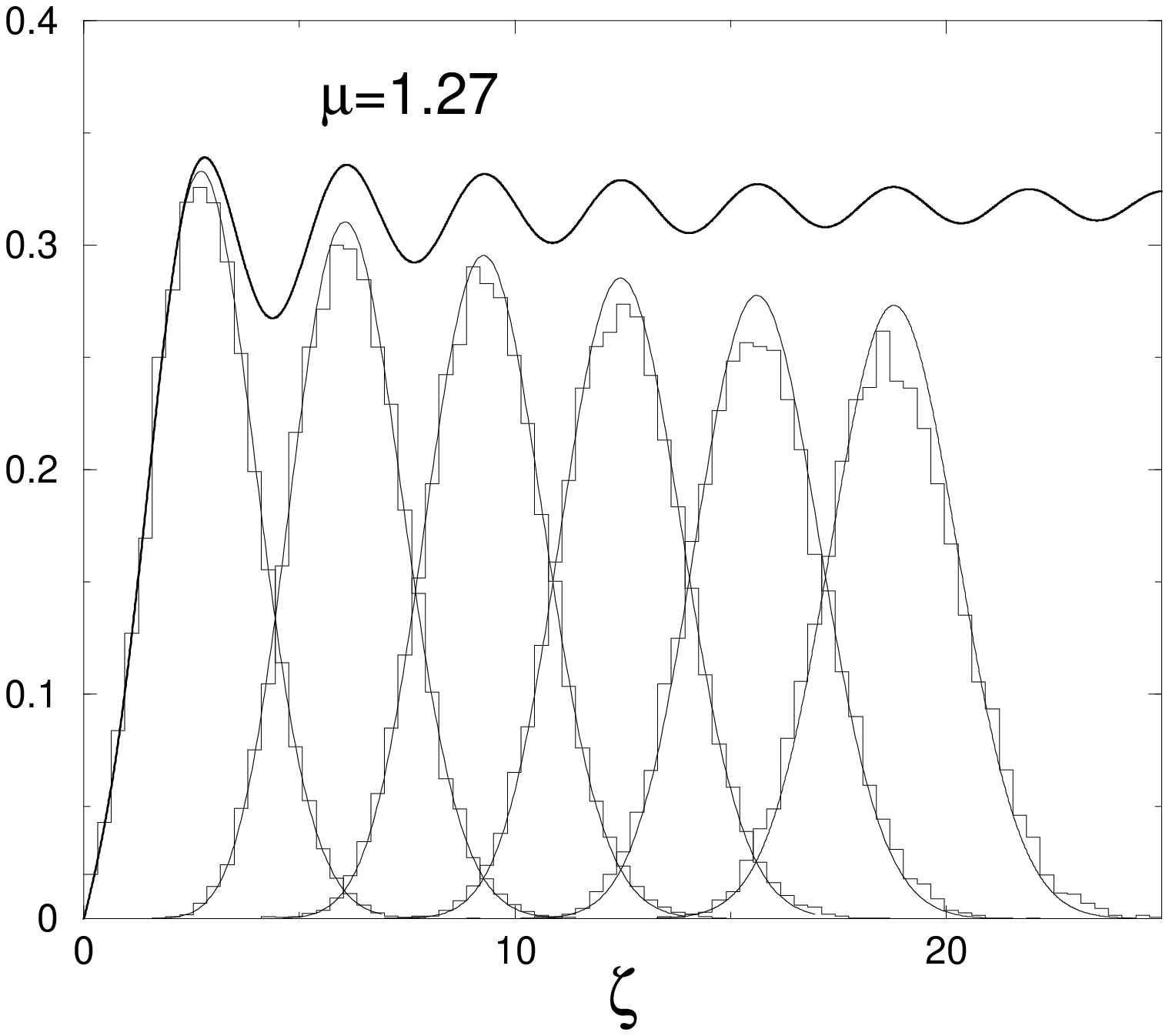}
\end{tabular}
\caption{The smallest eigenvalue distributions of the same data as in 
Fig.~\ref{fig:dyn_rho}. Agreement holds also at individual distributions.
The sum of the individual eigenvalues, the microscopic spectral density,
is shown by the upper curve on all three plots.}
\label{fig:dyn_n_eig}
\end{figure}

\begin{figure}
\begin{center}
\begin{tabular}{c c}
\includegraphics[scale=0.45]{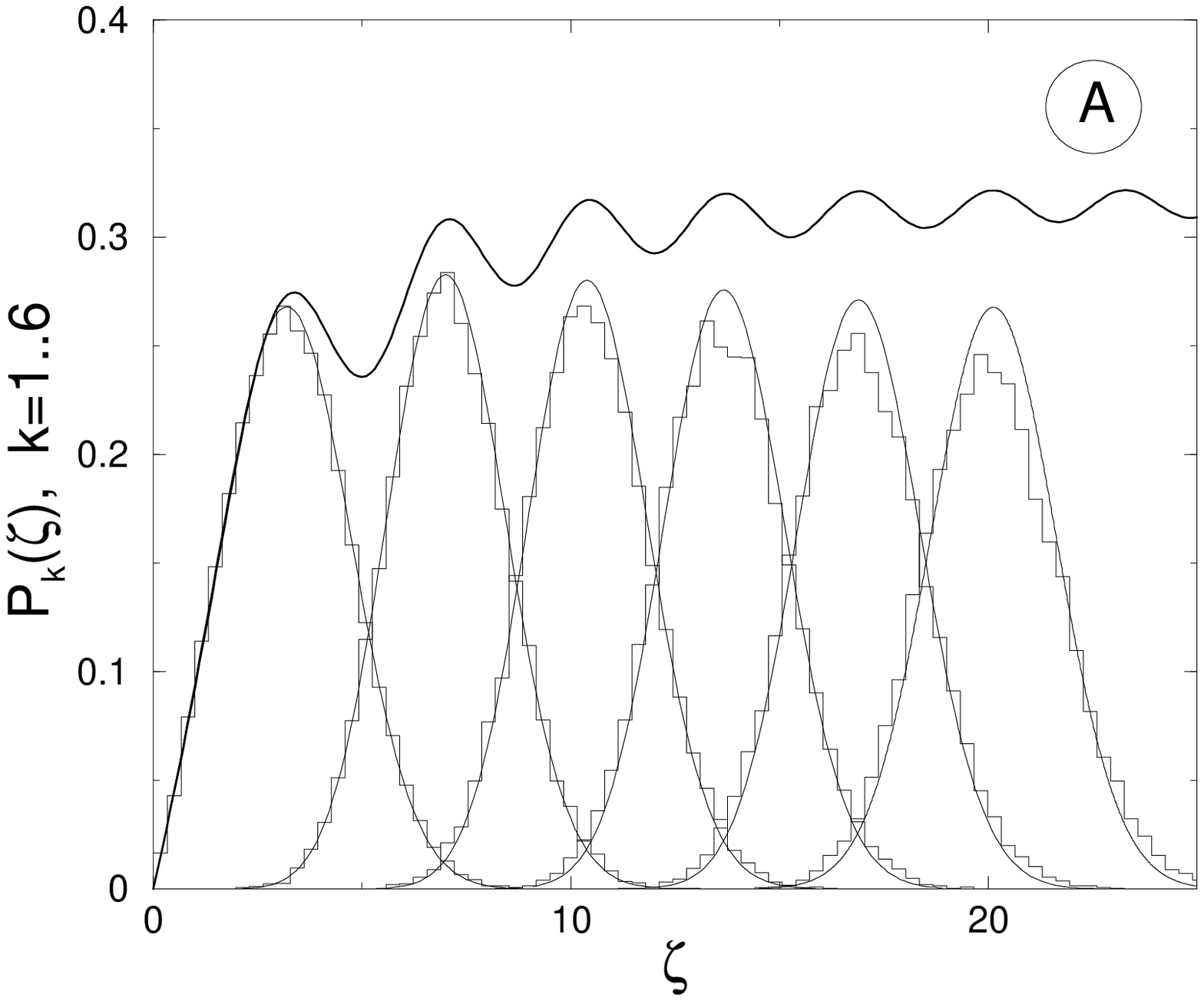}
&
\includegraphics[scale=0.45]{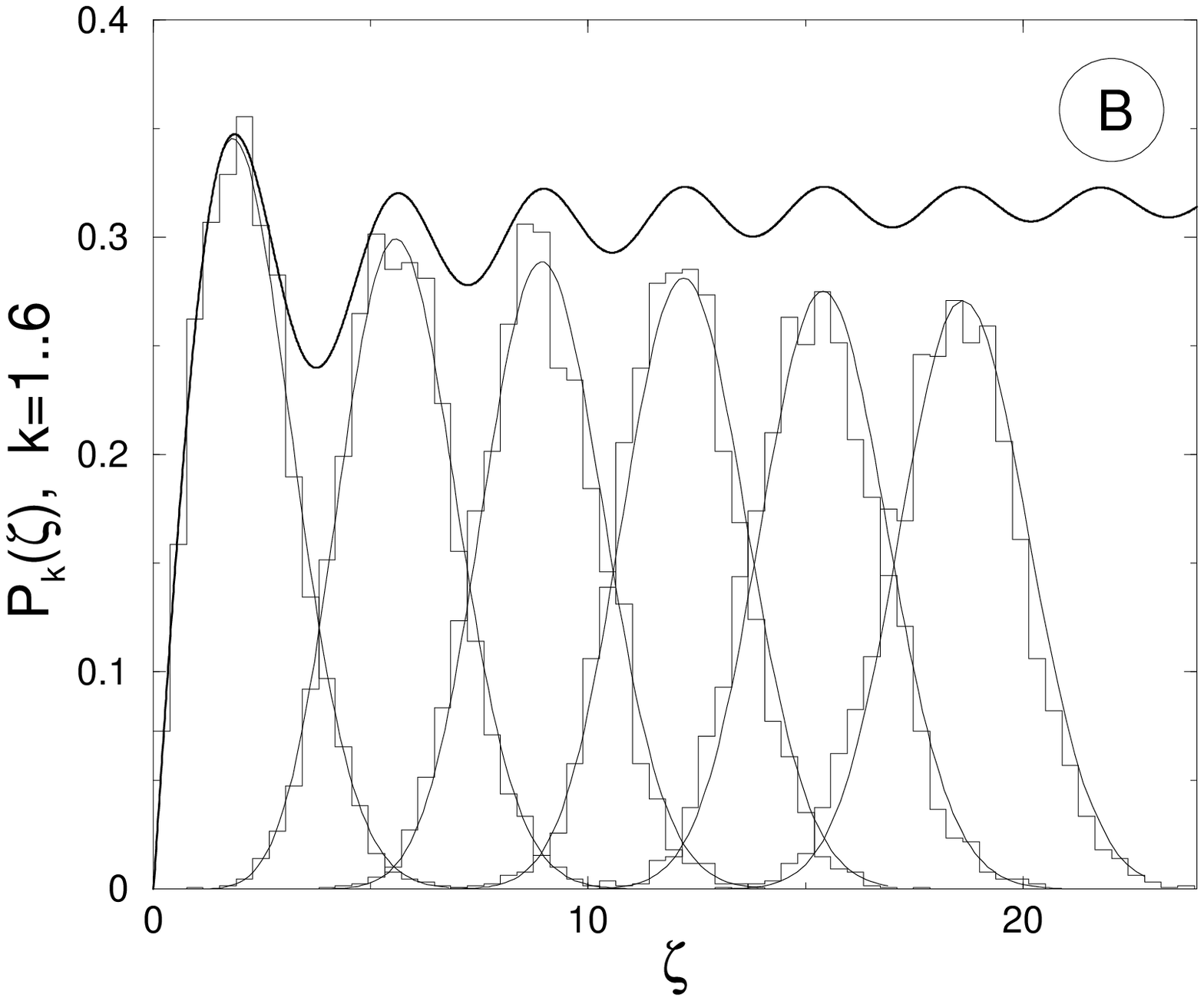}
\end{tabular}
\end{center}
\caption{The smallest eigenvalue distributions for ({\bf A}) the $N_f=2$
theory with degenerate masses on a $4^4$ lattice, and ({\bf B}) the $N_f=1$
theory on a $6^4$ lattice. Gauge couplings as shown in Table~\ref{tab:param}.
Again the spectral density corresponding to these $N_f$ and $\mu$ parameters
are shown as the upper curves.}
\label{fig:Nf2_V6}
\end{figure}

\noi In Fig.~\ref{fig:dyn_rho} we compare the spectral density of the
simulation data for different quark masses. $\Sigma$ was determined,
for each of the three ensembles, from a fit of the distribution of
the lowest eigenvalue. The values agreed within errors, and were
averaged to obtain the result in Table~\ref{tab:param}. This agreement
is of course as it should be: $\Sigma$ is the infinite-volume chiral 
condensate in the chiral limit $m \to 0$ (and in particular mass-independent).
For each mass we plot the
analytical result (\ref{eq:rhoS}) for that particular mass value
(rescaled according to $\mu = m\Sigma V$), and include the
curve for the limiting cases of the quenched result corresponding to
infinite mass quarks, as well as for one massless quark. We see that the 
data beautifully interpolate between the two limits and follow the
analytical curve almost perfectly. Looking at the individual eigenvalue
distributions in Fig.~\ref{fig:dyn_n_eig} we get a better view on how
far the agreement goes. It is clear that only the very first
eigenvalues match perfectly and that there is a very tiny discrepancy for
the last (of the 6) distributions due to the limited volume, but still
the agreement is impressive. We did a single run on a larger volume
($6^4$), and indeed the agreement is better, see Fig.~\ref{fig:Nf2_V6}
{\bf B}. Finally, data from a run corresponding to two flavors of
quarks are presented in Fig.~\ref{fig:Nf2_V6} {\bf A}, with the
same degree of agreement.

\noi
To conclude, we have found very precise agreement with the
analytical predictions for the distributions of Dirac operator eigenvalues
in the microscopic finite-volume scaling region. This is seen in
both quenched and unquenched simulations, and for the first time
there are now high-statistics comparisons between the analytical 
predictions for the Dirac operator spectrum of {\em massive} quarks
and lattice Monte Carlo data for gauge group SU(3). Our Monte Carlo
calculations have been restricted to staggered fermions, which are
insensitive to gauge field topology at the couplings
considered here. We have thus compared
with analytical predictions for the topological sector of charge $\nu=0$.
Perhaps the biggest surprise is that the previously seen agreement with
analytical predictions for the summed-up spectral density 
holds at a much more detailed level underneath. 
The individual smallest Dirac operator
eigenvalues are really completely locked up by gauge field topology, and, away
from the exact zero modes, by the coset $G/H$ of spontaneous chiral symmetry
breaking.

\vspace{1cm}
\noi
{\sc Acknowledgments:} The work of P.H.D. and K.R. has been partially 
supported by EU TMR grant no. ERBFMRXCT97-0122, and the work of U.M.H.
has been supported in part by DOE contracts DE-FG05-96ER40979 and
DE-FG02-97ER41022. P.H.D. and U.M.H. also 
acknowledge the financial support of NATO Science Collaborative Research
Grant no. CRG 971487.


\end{document}